\documentclass[12pt]{extarticle}

\usepackage[body={17.5cm, 22.5cm},right=2cm]{geometry}

\usepackage{booktabs}
\usepackage{caption}

\usepackage{color}
\usepackage{graphicx}
\usepackage{epsf}
\usepackage{graphicx,epsfig}
\usepackage{multirow}

\usepackage{amsmath, amsthm, amssymb}

\usepackage{epsfig}
\usepackage{cite}
\usepackage{color,colordvi}
\newcommand{\be}{\begin{eqnarray}}
\newcommand{\ee}{\end{eqnarray}}
\newcommand{\bi}{\begin{itemize}}
\newcommand{\ei}{\end{itemize}}

\newcounter{hran}

\makeatletter
\renewcommand\section{\@startsection {section}{1}{\z@}%
                               {-3.5ex \@plus -1ex \@minus -.2ex}%
                               {2.3ex \@plus.2ex}%
                               {\normalfont\large\bfseries}}
\makeatother

\numberwithin{equation}{section}
\begin{document}
\vspace{5mm}
\vspace{0.5cm}

\vspace{5cm}

\begin{center}

\def\thefootnote{\fnsymbol{footnote}}

{\large \bf 
  Decoupling Limits of sGoldstino Modes in Global and Local 
  \vskip.1in
  Supersymmetry
  %
}
\\[1.5cm]
{\large  Fotis Farakos and Alex Kehagias }
\\[0.5cm]

\vspace{.3cm}
{\normalsize {\it  Physics Division, National Technical University of Athens, \\15780 Zografou Campus, Athens, Greece}}\\

\vspace{.3cm}
{\normalsize { E-mail: fotisf@mail.ntua.gr,  kehagias@central.ntua.gr }}


\end{center}

\vspace{3cm}

\hrule \vspace{0.3cm}
\small  \noindent \textbf{Abstract} 
\noindent

We study the decoupling limit of  a superheavy sgoldstino field in spontaneously broken ${\cal{N}}=1$ supergravity.
 Our approach is based on K\"ahler superspace, 
which, among others, allows direct formulation of ${\cal N}=1$ supergravity in the Einstein frame
and correct identifications of mass parameters. 
Allowing for a non-renormalizable K\"ahler potential in the hidden sector, 
the decoupling limit of a superheavy sgoldstino is identified with an infinite negative K\"ahler curvature.
Constraints that lead to non-linear realizations of supersymmetry emerge as consequence of the 
equations of motion of the goldstino superfield when considering the decoupling limit.
Finally, by employing superspace Bianchi identities, 
we identify  the real chiral superfield, which will be
the superconformal symmetry breaking chiral superfield 
that enters the conservation of the Ferrra-Zumino multiplet in the field theory limit of ${\cal N}=1 $ supergravity.

\vspace{0.5cm}  \hrule
\vskip 1cm

\def\thefootnote{\arabic{footnote}}
\setcounter{footnote}{0}


\newpage



\baselineskip= 19pt



\def\ls{\left[}
\def\rs{\right]}
\def\lc{\left\{}
\def\rc{\right\}}

\section{Introduction}

Supersymmetry is one of the most appealing cadidates for 
new physics. It has not been observed so far and  
thus, it should  be broken at some high energy scale if it is realised at all.
However,  supersymmetry breaking is not an easy task. 
In the MSSM for example,  supersymmetry breaking is employed by  introducing 
soft breaking terms.
These terms are {\it ad hoc} masses for the superpartners of the SM particles, which nevertheless do not
spoil the 
UV properties of the theory. 
In fact the MSSM includes all these soft breaking terms and one has to fit them into the observations.
From a more theoretical point of view, the origin of  these soft terms should be explored.
The common lore is that supersymmetry should be broken in a sector of the theory, not directly connected to the SM particles,
the hidden sector.
{  For a review on soft terms, and other supersymmetry breaking mediation scenarios we refer to 
\cite{Freedman:2012zz,Weinberg:2000cr,Martin:1997ns}.}

Whatever the nature of the mediation, 
the hidden sector should be studied on its own right.
If it is a chiral multiplet that breaks supersymmetry, 
its highest component $F$ will acquire a non-vanishing {\it vev}.
There is a number of different scenarios for the origin of the supersymmetry 
breaking\cite{Martin:1997ns,Freedman:2012zz}.
Let us note that higher derivative operators \cite{Cecotti:1986jy,Koehn:2012ar,Farakos:2012qu,jeanluc}
may play an important role in 
  hidden sector supersymmetry breaking.  
One of the most efficient methods for studying the phenomenology of the hidden sector 
is through the dynamics of the goldstino\cite{Volkov:1973ix,Ivanov:1978mx,Rocek:1978nb,Lindstrom:1979kq,uematsu,ho-kim,Casalbuoni:1988xh,luty,
lee,brignole,antoniadis1,antoniadis2,bagger,seiberg1,kuz0,kuz00}. The latter is 
the fermionic component of the superfield that breaks supersymmetry.
If the supersymmetry breaking scale is low, 
goldstino dynamics become increasingly important for low energy phenomenology 
\cite{pheno1,pheno2,pheno3,pheno4,pheno5,AG1,AG2,AG3,Komargodski:2009rz,pantelis,Antoniadis:2012zz,Farakos:2012fm}.
In fact, if the  SUSY breaking scale $\sqrt{f}$ is low with respect to Planck mass $M_P$ 
($\sqrt{f}\ll M_P$) 
as in gauge mediation, transverse gravitino couplings are of order $M_P^{-1}$ and therefore 
are suppressed with respect to longitudinal gravitino couplings, which are of order $f^{-1/2}$. 
In this case, in the gravity decoupling limit, only the longitudinal gravitino component, 
i.e., the goldstino survives. Moreover, the highest component of the superfield to which the goldstino belongs, 
acquires a vev and breaks spontaneous the supersymmetry giving also mass to the sgoldstino (goldstino's superpartner).
Therefore, at low energies,
 supersymmetry is spontaneous broken and after decoupling  the sgoldstino (by making the latter superheavy)
 we are left with only the goldstino in the spectrum and  a non-linear realised SUSY. 
In the case of local supersymmetry,  non-linear realizations are less studied in the supergravity context
\cite{Lindstrom:1979kq,Gates:1983nr,deAlwis:2012aa}.

Recently new methods have been proposed in order to study goldstino couplings, 
and MSSM extensions that incorporate them have been constructed 
\cite{Komargodski:2009rz,pantelis,Dudas:2011kt,Antoniadis:2011xi,Antoniadis:2012zz,Farakos:2012fm,Antoniadis:2012ck,Dudas:2012fa}.
All this framework is based on the idea of constrained superfields\cite{Casalbuoni:1988xh,Rocek:1978nb,Lindstrom:1979kq} that introduce a  
non-linear supersymmetry representation for the goldstino when its  massive scalar superpartner is heavy 
and can be integrated out.
Moreover, when one studies physics much lower than the MSSM soft masses scale, 
non-linear supersymmetry is realized on the SM particles as well, 
via the appropriate constraints. 
The constraint that enforces a non-linear supersymmetry realization for the goldstino reads
\be
\label{intr}
\Phi_{NL}^2=0.
\ee
In addition, it has been proven in \cite{Komargodski:2009rz}   that in fact $\Phi_{NL}$ is proportional 
in the IR limit to the chiral superfield $X$ that sources the violation of the conservation of the 
Ferrara-Zumino supercurrent $J_{\alpha\dot{\alpha}}$ \cite{Ferrara:1974pz,CL} 
\be
\bar{D}^{\dot{\alpha}}J_{\alpha\dot{\alpha}}=D_\alpha X. \label{fz1}
\ee
 We extend this to the case of ${\cal{N}}=1$ supergravity by   identifying the  superfield,
 which turns out to be the chiral superfield $X$ of (\ref{fz1}) in the
 gravity decoupling limit. Here, the conservation of the Ferrara-Zumino multiplet $J_{\alpha\dot{\alpha}}$ in 
 (\ref{fz1}) is replaced by the consistency conditions of the Bianchi identities \cite{Binetruy:2000zx}
\be
\label{bianchicondition}
X_{\alpha}= {\cal{D}}_\alpha {\cal{R}} - \bar{{\cal{D}}}^{\dot{\alpha}}G_{\alpha \dot{\alpha}} 
\ee
where $G_{\alpha\dot{\alpha}}$ and ${\cal{R}}$ are the usual supergravity superfields 
and $X_{\alpha}=  -\frac{1}{8} (\bar{{\cal{D}}}^2 - 8 {\cal{R}} ) {\cal{D}}_\alpha K$ is the matter sector contribution.


\section{Supergravity in Einstein frame}

In the standard ${\cal{N}}=1$ superspace forlmulation 
of supergravity,  one is forced to perform a Weyl rescaling to the action in order to write the theory
in the Einstein frame. Here, we should  write the superspace  action directly
in the Einstein frame since  we want 
to correctly identify the masses to be send to infinity. 
 This will  provide the superfield equations of motion in the correct frame as well.
The appropriate framework for this is the K\"ahler superspace formalism which we will briefly present below. For a
detailed description, one may consult for example \cite{Binetruy:2000zx,Binetruy:1987qw,Grimm:1990fp}.   
An alternative method would be  a super-Weyl invariant reformulation of the old minimal 
formulation for N=1 SUGRA \cite{kuz1}.


In the conventional superspace approach to supergravity, 
the Lagrangian describing gravity coupled to matter would be (ignoring superpotential for the moment)
\be
\label{F1-Sugra}
{\cal{L}}_{\rm F}=
\int d^2 \Theta 
2 {\cal{E}} 
 \left\{\frac{3}{8} 
(\bar{\cal{D}}\bar{\cal{D}}-8{\cal{R}}) e^{-\frac{1}{3} K(\Phi,\bar{\Phi})}\right\} +h.c.
\ee
where $2 {\cal{E}}$ is the superspace chiral density and the new $\Theta$ variables span
only the chiral superspace. An equivalent way to write the action (\ref{F1-Sugra}) is 
\be
\label{D-Sugra}
{\cal{L}}_{\rm D}=
-3  \int d^4 \theta 
E
e^{-\frac{1}{3} K(\Phi,\bar{\Phi})} ,
\ee
where now $E$ is the full superspace density and $\theta$ are to be integrated over the full superspace.
Both actions (\ref{F1-Sugra},\ref{D-Sugra}) can equavalently be used in order to build invariant theories in superspace. 
Note that $ {\cal{E}}$ and $E$, both  have the vierbein determinant in their lowest component.
As usual $\cal{R}$ represents the supergravity chiral superfield which contains the Ricci scalar in its highest component.
Direct calculation of  (\ref{D-Sugra}) in  component form shows that the theory is 
actually expressed in an unconventional Jordan frame.
Of course a Weyl rescaling may be performed in order to bring the theory in 
the standard Einstein frame. 
Nevertheless, it is possible to perform this rescaling at the superspace level by considering
\be
\label{resc}
\nonumber
{E'}_{M}^{\ a} &=& e^{-\frac{1}{6} K(\Phi,\bar{\Phi})} E_{M}^{a}\, , ~~~
{E'}_{M}^{\ \alpha} = e^{-\frac{1}{12} K(\Phi,\bar{\Phi})} 
\ls E_{M}^{\ \alpha} - \frac{i}{12} E_{M}^{b} (\epsilon\sigma_b)_{\ \dot{\alpha}}^{\alpha} \bar{\cal{D}}^{\dot{\alpha}} K(\Phi,\bar{\Phi})  \rs ,
\\
\nonumber
{E'}_{M\dot{\alpha}} &=& e^{-\frac{1}{12} K(\Phi,\bar{\Phi})} 
\ls E_{M\dot{\alpha}} - \frac{i}{12} E_{M}^{b} (\epsilon\bar{\sigma}_b)_{\dot{\alpha}}^{\ \alpha} {\cal{D}}_{\alpha} K(\Phi,\bar{\Phi})  \rs
\ee
where $E_{A}^{M}$ is the superspace frame, containing the gravitino and the vierbein in the appropriate lowest components.
This redefinition will change the structure of the whole superspace including the Bianchi identity solutions and the superspace derivatives.
Most importantly, the superspace geometry will receive contributions at the same time from the matter and supergravity fields in a unified way.
The Lagrangian (\ref{D-Sugra}) now becomes in the new superspace frame (erasing the primes for convenience) 
\be
\label{D'-Sugra}
{\cal{L}}_{\rm Dnew}= -3 \int d^4 \theta E .
\ee
This form  now contains the properly normalized supergravity action coupled to matter. 
The interested reader should consult an extensive review on the subject \cite{Binetruy:2000zx}.
Since we also wish to include a superpotential, the appropriate contribution will be given by
adding to (\ref{D'-Sugra}) the appropriatelly rescaled superpotential $W$ so that the full  Lagrangian
will be given by 
\be
\label{P-Sugra}
{\cal{L}}_{superpotential}=  -3 \int d^4 \theta E\, +\, \left\{\int d^4 \theta \frac{E}{2{\cal{R}}} e^{K/2} W  + h.c.
\right\} .
\ee
In this new framework, K\"ahler transformations, generated by holomorphic functions $F$,  
are expressed as field dependent transformations gauged 
by a composite $U_{K}(1)$ vector $B_A$.
The respective charge now is referred to as ``chiral weight'' 
and a superfield $\varPhi$ of chiral weight $w(\varPhi)$ transforms as
\be
\label{w}
\varPhi \rightarrow \varPhi e^{-\frac{i}{2}w(\varPhi) \text{Im} {\cal{F}}} .
\ee
Gauge covariant superspace derivatives are defined as
\be
\label{d}
{\cal{D}}_A\varPhi = E_{A}^{\ M} \partial_M \varPhi  +w(\varPhi)  B_A \varPhi 
\ee
where the composite connection superfields are
\be
\label{A}
\nonumber
B_{\alpha}&=&\frac{1}{4}{\cal{D}}_{\alpha} K\, , ~~~
\bar{B}^{\dot{\alpha}}=-\frac{1}{4}\bar{{\cal{D}}}^{\dot{\alpha}} K
\\
\nonumber
B_a&=&\frac{1}{4}(\partial_iK) {\cal{D}}_a \Phi^i - \frac{1}{4}
 (\partial_{\bar{j}}K)
{\cal{D}}_a 
\bar{\Phi}^{\bar{j}} 
+\frac{3i}{2} {\cal{G}}_a +\frac{i}{8} g_{i \bar{j}} \bar{\sigma}^{\dot{\alpha}\alpha} 
( {\cal{D}}_{\alpha} \Phi^i ) \bar{{\cal{D}}}_{\dot{\alpha}} \bar{\Phi}^{\bar{j}}  .
\ee
All component fields are understood to be defined appropriately via projection as usual 
but now with the use of these K\"ahler-superspace derivatives.
It turns out that the invariant Lagrangian containing both (\ref{D'-Sugra}) and (\ref{P-Sugra}) depends only on the generalized K\"ahler potential
\be
\label{G}
e^G=e^{K(\Phi,\bar{\Phi})}W(\Phi)\bar{W}(\bar{\Phi}).
\ee
By taking into account the chiral weights of the gravity sector and performing a K\"ahler transformation with parameter ${\cal{F}}=\text{ln}W$, 
we find that the final expression for the most general coupling of matter to supergravity is
\be 
\label{K-Sugra}
{\cal{L}}= \int d^4 \theta E \ls -3 + \frac{1}{2{\cal{R}}} e^{\frac{G}{2}} + \frac{1}{2\bar{{\cal{R}}}} e^{\frac{G}{2}}  \rs.
\ee

It sould be stressed that this form of the action is completely equivalent to the standard ${\cal{N}}=1$ superspace 
formulation (\ref{F1-Sugra}) to which is related by appropriate redefinitions of the superspace frames.

\section{Sgoldstino decoupling} 

We are interested in those classes of models where 
the sgoldstino may  become superheavy and decouples from the spectrum. 
In this case, it  plays no role in the low energy effective theory, 
and its dynamics can be integrated out by its equations of motion.
Essentialy, in order to be able to decouple consistently the sgoldstino degrees of freedom, 
one has to 
\begin{enumerate}
\item consider the sgoldstino mass as the heavier scale in the problem, and
\item find consistent solutions for the equations of motion in that limit.
\end{enumerate}
This is equivalent to taking the limit of infinitely heavy sgolstino and integrate its equations of motion,
if possible, in this limit.  
This work has been done in component form earlier \cite{Casalbuoni:1988xh} 
and extended recently\cite{Antoniadis:2011xi,Dudas:2011kt}. We will implement the above procedure 
in superspace, where as we will see it is quite straightforward.

To study sgoldstino decoupling in supergravity, it is helpful to consider the corresponding decoupling in 
global supersymmetry. 

\subsection{Sgoldstino decoupling in global supersymmetry}
The most general single chiral globally supersymmetric superfield Lagrangian is given by
\be
\label{DF}
{\cal{L}}=\int d^4 \theta \, K(\Phi,\bar{\Phi}) \,+
\,\left\{ \int d^2 \theta \, W(\Phi) + h.c. \right\}
\ee
where, $K(\Phi,\bar{\Phi})$ is the K\"ahler potential, a hermitian function of the chiral superfield, 
and $W(\Phi)$ is the superpotential, a holomorphic function of the chiral superfield.
From the above action, the superspace equations of motion 
\be
\label{SF-EOM}
-\frac{1}{4} \bar{D}\bar{D} K_\Phi +  W_\Phi = 0,
\ee
with $K_\Phi=\partial_\Phi K,~W_\Phi=\partial_\Phi W$ easily follow. 
For a general, non-renormalizable supersymmetric model where supersymmetry is spontaneously broken, 
the supertrace mass formula reads \cite{Gates:1983nr}
\be
\label{susy-trace}
\text{Str} M^2 = \sum_{J} (-1)^{2J} (2J+1) M_{J}^{2} = -2  R_{A\bar{A}}  f \bar{f}
\ee
where $f=\langle F\rangle$ and
$R_{A\bar{A}}~~ (=g^{A\bar{A}}R_{A\bar{A}A\bar{A}} )
$
is the Ricci tensor of the scalar K\"ahler manifold evaluated at the vacuum expactation values of the scalars. 
Eq.(\ref{susy-trace}) describes the mass splitting between the components of the supermultiplet.
In the case of a single chiral superfield we are discussing, since the goldstino is always massless, 
the supertrace of the goldstino multiplet is just  the square of  the sgoldstino mass
\be
M_{\rm sg}^2=- R_{A\bar{A}} f\bar{f} \label{msg}
\ee
We see that necessarily the scalar manifold should be a space of negative curvature in order to have non-tachyonic 
scalar excitations.  In addition,
the limit of the infinitely heavy sgoldstino  
\be
\label{limit1}
2M_{sg}^2= \text{Str} M^2 \rightarrow \infty \,  ~~~\mbox{or}~~~
R_{A\bar{A}A\bar{A}} \rightarrow -\infty.
\ee
Since 
\be
R_{A\bar{A}A\bar{A}}=\partial_{\bar{A}}\partial_A\partial_{\bar{A}}\partial_A K-
\partial_{\bar{A}}\partial_A\partial_{\bar{A}} K\partial_A\partial_A\partial^A K\, , 
\ee
in normal coordinates for the K\"ahler space in which $g_{A\bar{A}}=\delta_{A\bar{A}}$ and 
$\partial_i\partial_j\partial_k K=0$ (for any $i,j=A,\bar{A}$), we have that the infinitely heavy sgoldstino 
is obtained in the limit 
\be
-\partial_{\bar{A}}\partial_A\partial_{\bar{A}}\partial_A K \to \infty
\ee
By assuming that the vacuum expectation value of $A=\Phi\big{|}$ vanish\footnote{if not we may shift 
appropriately $A$ 
so that  $\langle A\rangle=0$}, the general form of the K\"ahler potential 
\be
\label{K}
K(\Phi,\bar{\Phi})=\sum_{mn} c_{mn}  \Phi^m\bar{\Phi}^n
\ee
 will have the following expansion  in normal coordinates
\be
\label{Kahler1'}
K(\Phi,\bar{\Phi})=\Phi\bar{\Phi}+c_{22} \bar{\Phi}^2 \Phi^2 +\cdots\ \ \ 
\ee
It is easy to see that in fact 
\be
c_{22}=\frac{1}{4}R_{A\bar{A}A\bar{A}}=\frac{1}{4}R_{A\bar{A}}
\ee
in normal coordinates. By using then 
(\ref{susy-trace},\ref{limit1}), we get that the K\"ahler potential may be expressed in terms of 
the sgoldstino mass as
\be
\label{Kahler1}
K(\Phi,\bar{\Phi})=\Phi\bar{\Phi}-\frac{M_{sg}^2}{4 |f|^2} \bar{\Phi}^2 \Phi^2 +\cdots\ \ \ 
\ee
where the dots stands for $M_{sg}$-independed terms and $f=<F>$ is the vev of the auxiliary 
field in the chiral multiplet. 
From the superspace equations of motion (\ref{SF-EOM}), 
one can easily isolate the contribution proportional to $M_{sg}^2$. Indeed, 
(\ref{SF-EOM}) is  written as 
\be
 \frac{M_{sg}^2}{4|f|^2} \Phi \bar{D}\bar{D} \bar{\Phi}^2~+~\Big{(}
 M_{sg}\!\!-\!{\rm independed\, \,  terms}\Big{)} = 0.
\ee
Therefore, in the  $M_{sg}\to\infty$ limit, 
the $M_{sg}$-dependent part of the
 field equations 
is turned 
into  
the superspace constraint
\be
\label{constr1}
\Phi \bar{D}\bar{D} \bar{\Phi}^2=0.
\ee

To explicitly solve (\ref{constr1}), we note that it leads  to three component equations
\be
\label{const3}
\Phi\bar{D}\bar{D} \bar{\Phi}^2 |= 0,\ \ 
D_{\alpha}(\Phi\bar{D}\bar{D} \bar{\Phi}^2 )|= 0,\ \ 
DD(\Phi\bar{D}\bar{D} \bar{\Phi}^2 )|= 0.
\ee
The non-trivial solution to the above equations is \cite{Rocek:1978nb,Komargodski:2009rz} 
\be
\label{constsol1}
\Phi_{NL}=\frac{\chi \chi}{2F}+\sqrt{2}\theta \chi+\theta^2 F
\ee
which can be easily checked that it satisfies
\be
\label{const4}
\Phi_{NL}^2 =0. 
\ee
As a result, the sgoldstino can be safely decoupled in the $M_{sg}\to \infty$ limit as long as 
$\Phi$ satisfies (\ref{constr1}), or equivalently (\ref{const4}). 

\subsection{Sgoldstino decoupling in supergravity}
As in the case of global supersymmetry, we are interested in the equations of motion and 
the mass supertrace.
The superfield equations of motion as follow from the action (\ref{K-Sugra}) are \cite{Binetruy:1987qw}
\be
\label{EOM-sugra1}
&&{\cal{R}}=\frac{1}{2}e^{\frac{G}{2}},
\\
\label{EOM-sugra2}
&&{\cal G}_a+\frac{1}{8}G_{\Phi\bar{\Phi}}\bar{\sigma}^{\dot{\alpha}\alpha}_{a} {\cal{D}}_{\alpha} \Phi  \bar{\cal{D}}_{\dot{\alpha}} \bar{\Phi}=0,
\\
\label{EOM-sugra3}
&&(\bar{\cal{D}}\bar{\cal{D}}-8{\cal{R}}) {G}_\Phi=0.
\ee
On the other hand, for a general supergravity model with only one chiral multiplet the supertrace is given by \cite{Wess:1992cp}
\be
\label{sugra-trace}
\text{Str} M^2 = -2 R_{A\bar{A}}  f\bar{f},
\ee
which means that in the limit of infinite negative K\"ahler curvature the sgoldstino will become superheavy 
and can consistently be integrated out. Indeed, (\ref{sugra-trace}) is explicitly written as
\be
M_{sg}^2=2m_{3/2}^2- R_{A\bar{A}}  f\bar{f}.
\ee
Therefore, for finite gravitino mass $m_{3/2}$,  the infinite curvature limit 
\be
\label{limit2sugra}
R_{A\bar{A}A\bar{A}} \rightarrow -\infty
\ee
is equivalent to superheavy sgoldstinos. Again, in normal coordinates  
\be
R_{A\bar{A}A\bar{A}}=\partial_{\bar{A}}\partial_A\partial_{\bar{A}}\partial_A K=
\partial_{\bar{A}}\partial_A\partial_{\bar{A}}\partial_A G
\ee
and therefore with 
\be
G\supset\frac{2m_{3/2}^2-M_{sg}^2}{4|f|^2} \Phi^2\bar{\Phi}^2+\cdots
\ee
the decoupling limit we are after is again $M_{sg}^2\to \infty$. 
Taking into account that the K\"ahler curvature $M_{sg}^2/4|f|^2$ will dominate the equations of motion and
following the same reasoning as in the global supersymmetric case, we get  from (\ref{EOM-sugra3})
\be
\label{sugra-const1}
\Phi(\bar{\cal{D}}\bar{\cal{D}} -8{\cal{R}}) \bar{\Phi}^2 = 0.
\ee
This constraint is the curved superspace analogue of (\ref{constr1}). 
In order to solve it, we take into account that $\Phi(\bar{\cal{D}}\bar{\cal{D}} -8{\cal{R}}) \bar{\Phi}^2$ is a chiral superfield, 
and we will once again start from its lowest component, namely
\be
\label{const-lowest-sugra}
\Phi(\bar{\cal{D}}\bar{\cal{D}} -8{\cal{R}}) \bar{\Phi}^2| = 0.
\ee
This is written, 
for 
\be
\Phi=A+\sqrt{2}\Theta \chi+\Theta\Theta F\, , ~~~{\cal{R}}\Big{|}=-\frac{1}{6}M
\ee
as 
\be
\label{const-comp-sugra}
A M \bar{A}^2 -24 A \bar{A} \bar{F} + 12 A \bar{\chi} \bar{\chi} = 0.
\ee
This equation  has three solutions
\be
A_0=0, ~~~
A_1=\frac{\chi \chi}{2F},~~~
A_2= \frac{24 F}{M} -\frac{\chi\chi}{2F}.
\ee
The first solution $A_0$ is the trivial and we will not consider it. The second solution 
$A_1$ is the $\Phi^2=0$ we already encounter in the global susy case.  
The third solution  $A_3$ corresponds to $\Phi^2\ne0$ and can only be realized as long as
the auxiliary field of supergravity  $M$ is non vanishing ($M\neq0$).  However, from the
equation (\ref{EOM-sugra1}) we get 
\be
\label{split}
{\cal{R}}=\frac{1}{2}e^{\frac{G}{2}}=\frac{1}{2} e^{-\frac{M_{sg}^2}{8|f|^2} \Phi^2 \bar{\Phi}^2+\cdots},
\ee
where only the dominant term was explicitly written in the exponent in the right hand side. 
Now, in  the  
$M_{sg}^2 \rightarrow  \infty$ 
limit,
the right hand side goes to zero exponentialy fast so that for $\Phi^2\neq 0$
\be
{\cal{R}}=0  ~~~\mbox{for}~~~  M_{sg}^2\to \infty
\ee
Therefore also $M=-6 {\cal{R}}|=0$ and the third solution ($A_2$)
cannot  consistently be realized. As a result, 
the only solution to the constraint (\ref{sugra-const1}) is the 
$A_1=\frac{\chi \chi}{2F}$, or in other words 
the familiar
%
\be
\label{phiphi}
\Phi^2=0. 
\ee
This constraint leads to 
\be
\label{miracle}
e^{\frac{M_{sg}^2}{8|f|^2} \Phi^2 \bar{\Phi}^2}|_{\Phi^2=0}=1
\ee
and thus, the divergent part of (\ref{EOM-sugra1}) completely decouples!
Moreover, $\Phi^2=0$ also satisfies 
\be
\label{EOM-const2}
{\cal{D}}_{\alpha} \Phi  \bar{\cal{D}}_{\dot{\alpha}} \bar{\Phi}^2=0
\ee
which is   the field equation (\ref{EOM-sugra2}) in the $M_{sg}^2 \rightarrow \infty$ limit.
As a result, we have again arrived to the constraint (\ref{phiphi}) as the only viable and 
consistent condition for the decoupling of the sgoldstino.


\subsection{Supercurrent and sgoldstino decoupling} 

In order to discuss the relation of supersymmetry breaking to
conservation laws, let us explore the decoupling limit of the
supergravity sector. 
The supergravity  equations of motion (\ref{EOM-sugra1}) and (\ref{EOM-sugra2}) in
superspace, after 
 restoring dimensions with compensating powers of $M_P$ 
and returning to the K\"ahler frame where everything is expressed in terms of $K$ and $W$,
are written as 
\be
\label{EOM-sugra1k}
&&{\cal{R}}=\frac{1}{M_P^2}\frac{1}{2}We^{\frac{K}{2M_P^2}},
\\
\label{EOM-sugra2k}
&&{\cal
G}_a
+\frac{1}{M_P^2}\frac{1}{8}g_{i\bar{j}}\bar{\sigma}^{\dot{\alpha}\alpha}_{a} 
{\cal{D}}_{\alpha} \Phi^i  \bar{\cal{D}}_{\dot{\alpha}} \bar{\Phi}^{\bar{j}}=0.
\ee
Gravity decouples in the limit  $
M_P \rightarrow \infty$, 
and from (\ref{EOM-sugra1k}) and (\ref{EOM-sugra2k}) we have
\be
\label{Rdecoupled}
&&{\cal{R}} \rightarrow 0 ,~~~
{\cal G}_a  \rightarrow 0.
\ee
We note that this is the limit even when $W/M_P = \text{finite}$, which is
another possible  limit \cite{deAlwis:2012aa} for gauge
mediated SUSY breaking scenarios.
The fact that these supergravity superfields should vanish can be also 
understood from the algebra of supergravity when compared to supersymmetry.
For example, the global commutation relation (for $w(\Phi^i)=0$)
\be
\label{susy-com}
[\bar{D}_{\dot{\alpha}},D_a] \Phi^i=0,
\ee
in supergravity becomes
\be
\label{sugra-com}
[\bar{\cal D}_{\dot{\alpha}},{\cal D}_a] \Phi^i
=-i {\cal R} \sigma_{\alpha \dot{\alpha}}{\cal D}^{\alpha} \Phi^i
\ee
thus in order to recover the global supersymmetry algebra 
the superfield $\cal R$ should vanish.

Let us now derive the analog of the conservation equation of the Ferrara-Zumino multiplet
(\ref{fz1}) in curved superspace.
By using the consistency conditions of the Bianchi identities \cite{Binetruy:2000zx}
\be
\label{cc1}
X_{\alpha}= M_P^2 {\cal{D}}_\alpha {\cal{R}} -M_P^2 \bar{{\cal{D}}}^{\dot{\alpha}}G_{\alpha \dot{\alpha}} 
\ee
with 
\be
\label{cc2}
X_{\alpha}=  -\frac{1}{8} (\bar{{\cal{D}}}^2 - 8 {\cal{R}} ) {\cal{D}}_\alpha K
\ee
and the equations of motion, we find
\be
\label{curved-conservation}
 \bar{\cal D}^{\dot{\alpha}} {\cal J}_{\alpha \dot{\alpha}} 
=
{\cal D}_{\alpha} {\cal X}
-\frac{16}{3} {\cal R}{\cal D}_{\alpha} K 
+\frac{2}{3} {\cal G}_{\alpha \dot{\alpha}}\bar{\cal D}^{\dot{\alpha}}   K
\ee
with
\be
\label{J-sugra}
{\cal J}_{\alpha \dot{\alpha}}
= 2 g_{i\bar{j}}{\cal D}_{\alpha} \Phi^{i}  
\bar{\cal D}_{\dot{\alpha}}  \bar{\Phi}^{\bar{j}} 
-\frac{2}{3} [{\cal D}_{\alpha} ,\bar{\cal D}_{\dot{\alpha}}  ] K\, , ~~~~{\cal X} = 4 W e^{\frac{ K}{2M_P^2}} 
-\frac{1}{3} \bar{\cal D}\bar{\cal D} K.
\ee
The extra terms compared to (\ref{fz1}) arise due to 
commutation relations like (\ref{sugra-com}), 
and should vanish when supergravity is decoupled.

Now we take the decoupling limit of supergravity 
($M_P \rightarrow \infty$) 
with (${\cal R} \rightarrow 0,~ {\cal G}_a \rightarrow 0$) 
and find exactly the same formula as the global case.
As a final comment let us note that now, 
after the decoupling of supergravity,  
the superfield $X$ is
\be
\label{X-sugra2}
{\cal X} \rightarrow X= 4W -\frac{1}{3} \bar{D}\bar{ D} K .
\ee

\section{Conclusions}

In this work we explored the decoupling limit of  sgoldstinos in  spontaneously broken SUSY theories. 
This decoupling was implemented by considering large mass values for the sgoldstino 
(in fact the infinite mass limit).
We used superspace techniques as they allowed for a unified treatment of the spontaneous breaking of SUSY both in 
local and global supersymmetric cases. The motivation of this study was twofold: first
to check if the constraint superfield 
formalism employed in the global supersummetry still works in supergravity as well and second, to correctly 
identify in supergavity the 
chiral superfield that enters in the conservation of the Ferrara-Zumino multiplet and which 
accomodates the goldstino in global
supersymmetry. 

The way to approach these targets was to reformulate the goldstino dynamics in global supersymmetry 
but now in a language appropriate  for supergravity.
First we have identified the sgoldstino mass in both cases, 
and found the decoupling limit (supermassive sgoldstino) 
to be the limit of infinite negative K\"ahler curvature.
Then we impose this limit to the superfield equations of motion 
and in the case of supersymmetry we found the constraint $( \Phi \bar{D}^2\bar{\Phi}^2=0 )$
which is solved by $\Phi^2=0$ as expected.
In the case of supergravity, the super-covariant form of the more general constraint emerges, 
but again with the same single consistent solution.
Thus, the superspace constraint $\Phi^2=0$ for the goldstino, 
when the sgoldstino is supermassive, holds for supergravity as well.
However, we should mention a potential problem here. Namely, 
the expansion of the K\"ahler potential in (\ref{Kahler1}) is written in powers of $M_{sg}/f$, from where it follows that 
actually $M_{sg}\sim f/\Lambda$ where $\Lambda$ is the effective cutoff of the theory. The infinite sgoldstino mass
seems therefore to be  in conflict with the removal of the cutoff ($\Lambda\to\infty$), 
which is needed to identify the goldstino
superfield with the infrared limit of the superconformal symmetry breaking superfield that enters the Ferrara-Zumino 
current conservation. This issue is further complicated by the presence of extra light fields.
The problem has been pointed out in \cite{Antoniadis:2012ck} where  conditions
for the effective expansion 
of the supersymmetric Lagrangian in terms of the inverse cuttoff to not be in  conflict with a small sgoldstino  
mass $\sim f/\Lambda$ were given. Note that we have not faced this problem, as we have taken the formal infinite large sgoldstino
mass limit.





\section*{Acknowledgements}
We would like to thank Ulf Lindstr\"om for discussions and comments.
This research was implemented under the ``ARISTEIA" Action of the 
``Operational Programme Education and Lifelong Learning''
and is co-funded by the European 
Social Fund (ESF) and National Resources.


\end{document}